# Increasing Coverage of Indoor Localization Systems for EEE112 Support

Hamada Rizk
WRC, E-JUST, Egypt
and Tanta University, Egypt
Email: hamada.rizk@ejust.edu.eg

Moustafa Youssef
Dept. of Comp. and Sys. Eng.,
Alexandria University, Egypt
Email: moustafa@alexu.edu.eg

## Abstract

**Among many techniques for indoor localization, fingerprinting has been shown to provide a higher accuracy compared to the alternative techniques. Fingerprinting techniques require an initial calibration phase during which site surveyors visit virtually every location in the area of interest to manually collect the fingerprint data. However, this process is labour intensive, tedious, and needs to be repeated with any change in the environment.**

**In this work, we propose a technique for enhancing cellular-based indoor localization fingerprinting systems by automatically increasing the spatial density of the reference points. This can be achieved by generating synthetic measurements for virtually all points in the environment to cover inaccessible places.**

## Introduction

People spend most of their time indoors (e.g. shopping malls, public institutions, hotels, ... etc) [1]. Global positioning system (GPS) is unable to work indoors due to the absence of a line of sight to the reference satellites. Therefore, many research efforts are devoted towards transforming the way people navigate indoors in the same way as GPS. Many different techniques have been proposed in the literature which focus on leveraging already existing technology such as Wi-Fi [2], cellular [3][4] or Bluetooth [5] to enable localization services indoors. Based on the chosen infrastructure, fingerprinting techniques are harnessed to satisfy the accuracy constraint required for indoor localization.

Practically speaking, fingerprinting is a two-phase technique, consisting of an offline and an online phase. Received signals (i.e. fingerprints) are acquired at known points in the area of interest with some density (i.e. number of anchors per square meter), in the offline phase. Subsequently, the collected fingerprints are utilized to construct a localization model. This model can later be queried in order to estimate the user position in the online/tracking phase.

Fingerprinting-based systems are widely adopted as they are shown to yield fine-grained positioning. However, the deployment cost of such systems is prohibitive as it is time consuming and labour intensive. Moreover, their performance relies on the density of anchor points in the area of interest. State-of-the-art systems [3][4] rely on collecting signatures with high anchor point densities. Their accuracy drops significantly when they are used in settings with low anchor densities or containing inaccessible areas from which fingerprints were not taken. This gives rise to the challenge of providing precise and robust localization with minimal coverage of anchors in the area of interest.

In this paper, we introduce a method for increasing the anchor point density of cellular-based localization techniques used mainly for the Egyptian Enhanced Emergency Public Safety Service (EEE112). This will enable these techniques to perform better and to obtain a fine-grained accuracy. Specifically, instead of collecting signals at every point in the area of interest, we record fingerprints at a few sparse seed points. These fingerprints are then used to train a spatial interpolator for generating synthetic data of any spatial point in the environment. This helps to provide reasonably large training data without incurring high overhead on the user. As a result, the learning capability of the localization system is improved. We also provide an implementation for a fingerprint collector App on Android phones and adopt different localization techniques.

## Related work

This section presents a brief background on the current indoor localization approaches related to the proposed technique.

In fingerprinting-based approaches, a fingerprinting database usually called a radio map is constructed during the system offline/training time. This radio map consists of received signal strengths overheard from different transmitters in the area of interest as received at some anchor points. These transmitters can be access points or cell towers for WiFi or Cellular-based localization respectively. During the online tracking time, the overheard scan received at the mobile device of the user is matched against signatures of the signal at all locations to select the best match to the unknown location of the user.

The most common ways for implementing this approach are the probabilistic methods [8] [9] [10]. These methods aim at fitting the signal strength distribution at each location from each transmitter. In the tracking time, the location at which the probability of detecting this scan is maximized is selected to be user's position. Many variations of probabilistic fingerprinting-based indoor localization techniques have been devoted to improve the localization quality. In particular, taking the correlation between consecutive signal strength samples from the same signal source into account helps in enhancing the localization system performance [11]. For increasing the sensitivity of the system to small scale signal variations the system in [12] is proposed. To reduce the computational overhead, clustering locations based on their corresponding received signals that share the set of transmitters is employed in [13]. Another technique is proposed in [14] to study the multivariate probability distribution for location determination. Many recent method have been proposed, harnessing deep learning to obtain a fine-grained localization such as [6][23][24].

To avoid fingerprinting efforts, many systems [20] [21] have been proposed to characterize the relation between signal strength and distance using a propagation model. For example, the calculation the attenuation due to walls and other objects in the environment to estimate the location. Another method was proposed to leverage these propagation techniques to construct WiFi fingerprints [22].

Many systems over the years have been proposed to tackle the indoor localization problem based on the available sensors of modern smartphones [15][16][17][18][19]. Dead-reckoning has been used intensively in these systems, leveraging the smartphones' inertial sensors (e.g. magnetometer, accelerometer and gyroscope) to provide real-time tracking. However, the readings of such sensors are highly noisy and the accumulation of the noise are increasing rapidly leading to incorrect estimation of the user location. To handle error accumulation problem; Many methods have been proposed to reset the error using different techniques such as map-matching to the floor-plan [27][28], or leveraging physical and virtual landmarks [25][26].

## Methodology

Without a loss of generality, we assume a 2D area of interest where n cell towers can be heard. The offline stage is initialized by obtaining cell information – the cell tower ID and its corresponding received signal strength (RSS) - at few sparse seed points in the area of interest. This is carried out by the Fingerprint Collector App running on the phone. During each scan, the phone can receive signals from up to seven cell towers as reported in the GSM standards [6]. The scanned samples are then fed to a preprocessor for extending the feature set of each scan to include all heard cell towers in the area of interest among all scans. Consequently, the spatial generator module is initialized with 70% of the seed points and evaluated using the remaining 30%. We have tried several techniques as the core of this module. We found that the k-nearest neighbor regression (KNNR) yields the best performance in reconstructing the signals for the unknown point. This is due to its robustness to noisy data and outliers [7]. Therefore, it is adopted for this purpose. The output of the regressor is quantized to the RSS range of [0, 31] ASU, as the reading of RSS is reported as a discrete integer within this range by the cell phone APIs. With the use of this scheme, dense anchors can then be obtained at any point in the area of interest with minimal fingerprinting efforts. This can subsequently be used by any localization system which is shown to boost the localization accuracy.

## Evaluation and Results

We have implemented the proposed technique on different Android phones and evaluated it in a university building as applied to two localization systems [3][4]. Our results show that the proposed technique with a 2874% increase in coverage (with an increase of anchor point density from 0.39 point/m$^2$ to 11.49 point/m$^2$). We expect to see an improvement of 30% to 50% in the localization accuracy of the evaluated systems after applying this technique.

## ACKNOWLEDGMENT

This work has been supported in part by a grant from the Egyptian National Telecommunication Regulatory Authority (NTRA).